\documentclass[conference,letterpaper]{IEEEtran}

\usepackage[letterpaper, portrait, margin=0.5in]{geometry}

\usepackage{graphicx}

\usepackage{comment}
\usepackage{subcaption}
\usepackage{outlines}
\usepackage{xcolor}

\usepackage{etoolbox}
\usepackage{xcolor,colortbl}
\usepackage{multirow}
\usepackage{threeparttable}  
\usepackage[labelsep=space,justification=justified,singlelinecheck=false]{caption}
\usepackage{algpseudocode}

\usepackage[utf8]{inputenc} 
\usepackage[T1]{fontenc}
\usepackage{url}
\usepackage{ifthen}
\usepackage{cite}
\usepackage[cmex10]{amsmath}

\newcommand\NoThen{\renewcommand\algorithmicthen{}}

\begin{document}

\title{ A New Lossless Data Compression Algorithm Exploiting Positional Redundancy}
 
 \author{\parbox{4 in}{\centering Pranav Venkatram \\
 Email: pranavvenkatram@gmail.com
 }
\thanks{}
 }

\maketitle
\thispagestyle{empty}
\pagestyle{empty}

\begin{abstract}
A new run length encoding algorithm for lossless data compression that exploits positional redundancy by representing data in a two-dimensional model of concentric circles is presented. This visual transform enables detection of runs (each of a different character) in which runs need not be contiguous and hence, is a generalization of run length encoding.  Its advantages and drawbacks are characterized by comparing its performance with TurboRLE.
\end{abstract}

\section{Introduction}
Data compression involves the reduction of file sizes by exploiting redundancy in data i.e. repetition of characters. In lossless compression, algorithms are mainly of two types – Dictionary encoders e.g. Lempel-Ziv [1] family or Entropy encoders e.g. Huffman encoding [2]. These algorithms are conventional compressors which exploit high usage patterns and character distribution as types of redundancy respectively.\\

Traditionally, types of repetition such as positional redundancy are not usually taken advantage of. The exploitation of this particular type depends on characters repeating at predictable positions. It is generally accepted that most file types such as text have virtually no positional redundancy [1] as characters repeat at random positions. Therefore, positional redundancy is not considered significant to exploit and has not been thoroughly explored through dedicated compression methods.\\

This paper presents a lossless algorithm capable of leveraging on positional redundancy by overcoming the limitation of characters repeating at seemingly random positions in data. This is accomplished by representing said data in a two-dimensional structure as opposed to conventional methods which treat data as a one-dimensional stream.\\

The proposed method is compared with the TurboRLE algorithm in order to experimentally quantify its benefits. These tests are run on standard datasets such as the Silesia Compression Corpus, a dataset of files that encapsulate the typical data types used nowadays.

\section{Proposed Compression Algorithm}

\subsection{Concept and Analogy}
The proposed algorithm works by visualizing the input data in a two-dimensional model where each character is represented in a 2-D polar co-ordinate space, having theta $(\theta)$ and radius ($r$) value pairs with respect to an arbitrarily chosen reference axis.\\

For example, consider the input string S:\ THE PHONE BLAH\\

Step 1: S is parsed in order to split into substrings. For simplicity in this example, let us assume that S is split using the space character as a delimiter, yielding the substrings:

\begin{enumerate}
    \item THE,   $r$ = 1
    \item PHONE, $r$ = 2
    \item BLAH,  $r$ = 3
\end{enumerate}

\begin{figure}[h!]
    \centering
    \includegraphics[scale=0.4]{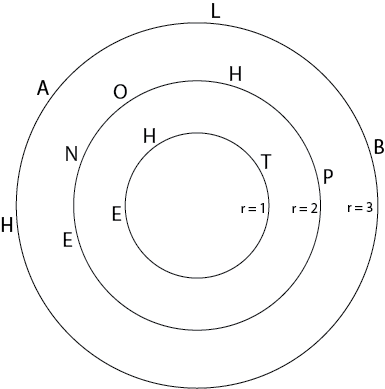}
    \caption{\centering{2-D Concentric circle representation of data}}
    \label{concentric_circle_2D}
\end{figure}

Step 2: Each substring will be considered a circle on which the characters are present. Therefore, all characters within the same substring have the same $r$ value but different $\theta$ values to denote their relative positions. Also, all characters on the first substring have a radius of $r$ = 1 as they are present on the first circle. Similarly, following substrings have a radius of one unit more, forming concentric circles. This model will hereafter be referred to as the Concentric Circles compression (CC) model which can be visualized in the form as seen in Fig. $\ref{concentric_circle_2D}$. Characters are represented anti-clockwise by convention.

This method focuses on the fact that each character’s position on its circle is relative to the characters occurring before and after it. Consider a letter having theta value $\theta$ and the characters before and after it have theta values $\theta_1$ and $\theta_2$ respectively. Since $\theta$ is relative, its value can be modified within the constraints of the following inequality, ensuring that the order of characters is not altered as in (1). 

\begin{equation}
    \theta_{1} < \theta < \theta_{2}
\end{equation}

This condition allows for the rotation of characters on their respective circles, allowing them to line up with the same character on the next concentric circle as their $\theta$ values will match.\\

\begin{figure}[t!]
    \centering
    \includegraphics[scale=0.38]{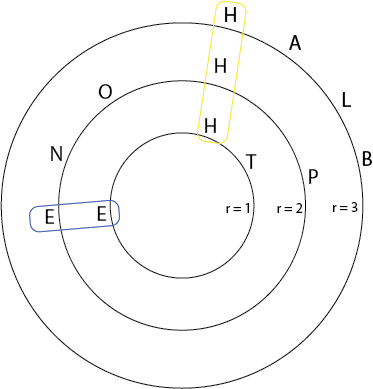}
    \caption{\centering{Transformed concentric circle structure}}
    \label{transformed_concentric_circle}
\end{figure}

Step 3: The letters H and E occur on consecutive concentric circles through which compression will take place. The position of letters will be altered by rotating them about the common centres of the circles such that they line up, yielding the altered structure in Fig. $\ref{transformed_concentric_circle}$. \newpage

Note that while the positions of each letter have changed, their relative positions remain the same, ensuring integrity in data. It is observed that the letter H repeats for $r$ = 1, 2, 3 and E repeats for $r$ = 1, 2. At this point, compression is achieved as instead of storing each $r$ value at which the letter occurs, only the start ($r_1$) and end ($r_2$) values need to be known. This is because it is implied in the concentric circle structure that the co-ordinates for all intermediate $r$ values at that particular $\theta$ value all store the same letter. For large datasets, the consecutive repetition of characters between circles can be much longer, improving compression. In practice, common letter occurrences are detected when substrings are parsed one at a time from $r$ = 1 (innermost circle) to the outermost circle as shown in Fig. $\ref{transformed_concentric_circle}$ for which $r$ = 3.\\

In Step 1, it was assumed that substrings would be formed using the space character as a fixed delimiter. The use of a delimiter is not the most optimal method because circles containing multiple occurrences of the same characters may exist. This reduces the data’s compressibility as compression in this method can only be achieved through consecutive character repetition between circles. Therefore, for correct substring generation, each character in each circle must be unique. Characters are added to a buffer from S until a new character being parsed is a pre-existing element of the buffer. At this point, the buffer contents become the latest substring and the next substring (new buffer) begins from the current character. Following this process, the string THEPHONEBLAH is split (spaces are excluded from the initial string for simplicity):

\begin{enumerate}
    \item THEP, $r$ = 1
    \item HONEBLA, $r$ = 2
    \item H, $r$ = 3
\end{enumerate}

\subsection{Exception Cases}
\begin{figure}[h!]
    \centering
    \includegraphics[scale=0.4]{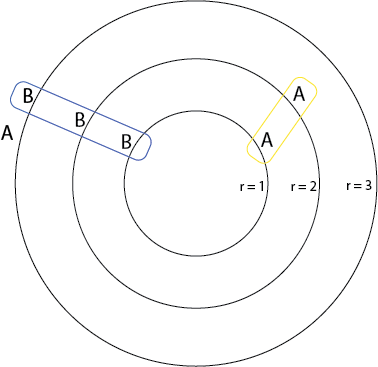}
    \caption{\centering{Spatial representation of example string}}
    \label{spatial_rep}
\end{figure}

In certain cases, it is not possible to compress characters. Consider the example string: ABABBA. It will be split into the following substrings and will be spatially represented as seen in Fig. $\ref{spatial_rep}$.
\begin{enumerate}
    \item AB
    \item AB
    \item BA
\end{enumerate}

Compression proceeds normally until the last circle. Here, the letter A cannot be lined up with the consecutive occurrences of A on the other circles as doing so would violate the previously stipulated condition for modifying $\theta$ values, thereby changing the order of letters. Since the letter B would have been parsed first, it will be given priority in terms of compression. This paradox (exception) case occurs when a seemingly common letter (A in this case) occurs before other common letter(s) (B in this case) on one circle and also occurs after the same common letter(s) on the next concentric circle. A more graphical explanation can be made: the straight lines on which common letters are present after rotation can never cross/overlap lest the order of letters change.\\

The two-dimensional nature of this algorithm allows for runs of non-contiguous characters (each of a different character) to be compressed, thereby generalizing Run Length Encoding to exploit positional redundancy. \\

\subsection{Practically Ensuring Compression}
In using CC, characters have associated $\theta$ and $r$ values. In order to optimize file size, neither $\theta$ nor $r$ values need to be stored for literals i.e. all non-compressible characters will remain as they are in the string and only the compressible ones will be removed and written to a separate file.\\

In the second file containing the compressed letters, only the starting $r$ ($r_1$) and ending $r$ ($r_2$) values are relevant. Hence, the repetition count/depth i.e. $r_2 -r_1$  and the $r_1$ values for repeating characters are stored. To ensure that this data is not too large, the $r_1$ values are delta encoded, denoted by the scheme in section $\ref{$r_1$ Delta Encoding Scheme}$. $\theta$ values only denote the relative positions of letters which is implicit in the stored and sequential output of the compressed letters. Hence, $\theta$ values do not need to be stored. In summary, this file contains serialized entries of three bytes: character, $r_1$, repetition count. Hereafter, this set of variables will be referred to as entries.\\

To know the position of compressed characters in the regeneration of uncompressed data during decompression, a bit flag prefix is assigned to each occurrence of a compressed character as well. Hence, a 0 bit indicates that the following character is a literal and a 1 bit in the output indicates that a compressed letter should fill that position. These bit flags allow for a one-dimensional string to be on par with the two-dimensional CC model although it does require additional data (an eighth of the input size in bytes) to be added to the compressed output. \\

The addition of bit flags causes a fixed overhead. In addition, when an entry containing a repetition count of 2 is stored, an overhead of 1 byte is incurred. This is because an entry is expressed as 3 bytes while only compressing 2 characters. Conversely, traditional RLE expresses an entry as 2 bytes, thus breaking even when compressing a run length of 2. Hence, entries of repetition count 2 in CC (hereafter referred to as redundant entries) must be removed.  \\ 

\subsection{Implementation of Proposed Algorithm}
An alphabet size of 256 (ASCII 8) is used for this investigation. \\

\subsubsection{$r_1$ Delta Encoding Scheme}
\label{$r_1$ Delta Encoding Scheme}
After encoding the input data, the compressed entries are parsed and their respective $r_1$ values are delta encoded with the $r_1$ of the node with the greatest $r_2$ value at that point in time i.e. $r_1$ + repetition count (hereafter referred to as a reference node). By convention, repetition count values are restricted to maximum of 127. This is essential to the delta encoding scheme and will be explained further on.\\
It can be inferred that all reference nodes cause an increase in $r$ value during compression as they have been created by triggering a break in the buffer string. This is because a character that is an element of the buffer has been encountered. 
This property is essential as entries occurring after the reference node in the serialized output will be within -127 to 128 units of $r$ (since repetition count can have a maximum value of 127) and hence their $r_1$ values can be expressed within a byte until such time a new reference node is encountered. 
The pseudocode of this is shown in \ref{delta encoding code}.   \\

\subsubsection{Proposed Data Structure to store Compressed Character Entries}
As stated before, the compressed characters and literal characters are separately stored. The compressor uses a hash table (collision resistance with separate chaining) with an embedded doubly linked list.\\

\begin{figure}[h!]
    \centering
    \includegraphics[scale=0.3]{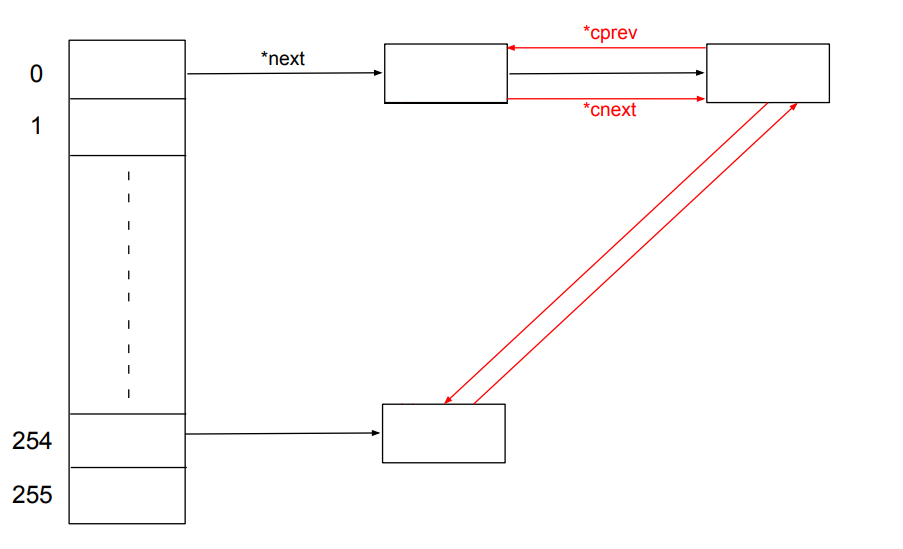}
    \caption{Hash table with separate chaining and an embedded doubly linked list}
    \label{hash_table}
\end{figure}

Each node contains the character, its $r_1$ value, its depth ($r_2 -r_1$), a pointer to the next node in the hash table as well as two other pointers which point to the previous and next node of the doubly linked list. This combined data structure is diagrammatically represented in Fig. $\ref{hash_table}$.

\begin{enumerate}
    \item *next points to the next node in the hash table due to separate chaining
    \item *cprev points to the previous node in the list of common letters with the previous relative theta value
    \item *cnext points to the next node in the list of common letters with the next relative theta value
\end{enumerate}

Since nodes are stored sequentially in the doubly linked list, theta values do not need to be stored. Each node’s position in the doubly linked list already indicates the character’s relative position ($\theta$ value). Thus, the combined use of a doubly linked list and a hash table with separate chaining allows data to be sorted as well as accessed in constant time.\\

The hash table has a size of 256 wherein every repeating character that is added in is stored at its ASCII value index at the head of the list in the hash table where the hash function returns the character’s ASCII value. Thus, every row of linked lists in the hash table store different entries of the same character.\\

Hash tables are assumed to search in O(1). However, due to the nature of this implementation, only the head of each singly linked list (most recent entry for the character) in the hash table needs to be searched, resulting in a guaranteed constant time search. This data structure is simply the proposed mechanism for use in code; other methodologies for storage can be used as well.\\

\subsubsection{Removal of Redundant Entries}
When the doubly linked list is parsed to compute the delta encoded $r_1$ value for entries, redundant entries can be uncompressed and deleted. However, if the redundant entry is a reference node, all following entries prior to the next reference node will compute their delta encoded $r_1$ value with respect to the previous reference node. There is a chance that this delta encoded value will be beyond the range -128 to 127 and hence cannot be stored losslessly in a single byte. Hence, redundant entries that are also reference nodes will only be uncompressed and deleted if all following entries prior to the next reference node can contain the new delta encoded $r_1$ value within the range -128 to 127. Other redundant entries can be uncompressed automatically.
The pseudocode of this is shown in \ref{removal of redundant entries}. \\

\subsubsection{File Handling}
In practice, the bit flags (from CC) are packed into sets of eight as the smallest unit of data a compiler can handle is a byte.
Practically, the two files containing literals and compressed characters respectively can be concatenated as long as a header of appropriate size is used to indicate where each file begins and ends along with their respective bit flags. Alternatively, an EOF code could be used.\\

\subsection{Pseudocode of Concentric Circles Compression}

\subsubsection{Encoding}
\label{delta encoding code} 
\label{removal of redundant entries}
\begin{algorithmic}[lines]
\State String buffer, prevstring, compressed output
\State Integer rvalue = 0
\State Integer greatestr1 = 0
\State Integer greatestdepth = 0\\
\While{not end of file}
\State Read next input character c

\NoThen
\If {c is an element of buffer}{
	\State Replace contents of prevstring with buffer’s
	\State Clear buffer
	\State Append c to buffer
	\State Increment rvalue by 1

   \NoThen
    \If {\parbox[t]{\dimexpr\linewidth-\algorithmicindent-\algorithmicindent\relax}{c is not present at a node in the hash table for which  $r_1 \le rvalue - 1 \le r_2$ AND no characters after c in prevstring have nodes in the hash table for which the same inequality holds true}}
    
        \State \parbox[t]{\dimexpr\linewidth-\algorithmicindent-\algorithmicindent\relax}{Create new node for c with $r_1 = rvalue-1$ and depth = 2, insert at the head of the singly linked list in the chained hash table and at the tail of the doubly linked list\\}
        
        \State \parbox[t]{\dimexpr\linewidth-\algorithmicindent-\algorithmicindent\relax}{Remove both occurrences of c from input data. Modify the bit flag for the first occurrence to 1 and add another bit flag at the current position of c with a value of 1\\}

   \NoThen
    \ElsIf {\parbox[t]{\dimexpr\linewidth-\algorithmicindent-\algorithmicindent\relax}{c is not present at a node in the hash table for \newline which $r_1\le rvalue-1\le r_2$ AND at least \newline one character after c in prevstring have nodes in \newline the hash table for     which the same inequality \newline holds true}}
    
        \State \parbox[t]{\dimexpr\linewidth-\algorithmicindent-\algorithmicindent\relax}{Create a new node for c with $r_1 = rvalue - 1$ and depth = 2, insert at the head of the singly linked list in the chained hash table and at the node before the tail in the doubly linked list\\}
    
        \State \parbox[t]{\dimexpr\linewidth-\algorithmicindent-\algorithmicindent\relax}{Remove both occurrences of c from input data. \newline Modify the bit flag for the first occurrence to 1 \newline and add another bit flag at the current position \newline of c with a value of 1}
        \\

    \Else
        \State \parbox[t]{\dimexpr\linewidth-\algorithmicindent-\algorithmicindent\relax}{Increment the depth value in c’s node by 1
        \\
    
        Remove the latest occurrence of c from the input \newline data and add a bit flag of value 1 at the current \newline position}
        \\
    \EndIf

    }
    
\\

\NoThen
\ElsIf{c is an element of prevstring}

    \State Append c to buffer\\
    
    \NoThen
    \If{\parbox[t]{\dimexpr\linewidth-\algorithmicindent-\algorithmicindent\relax}{a common letter occurs after c in prevstring AND \newline the same letter occurs before c in buffer}
    
    \\}
        \State \parbox[t]{\dimexpr\linewidth-\algorithmicindent-\algorithmicindent\relax}{ Do nothing as this is the exception/paradox case}
    \\
    \ElsIf {\parbox[t]{\dimexpr\linewidth-\algorithmicindent-\algorithmicindent\relax}{c is not present at a node in the hash table for \newline which  $r_1\le rvalue-1\le r_2$}
    \\
    \\}
        \State \parbox[t]{\dimexpr\linewidth-\algorithmicindent-\algorithmicindent\relax}{Create a new node for c with $r_1 = rvalue-1$ and depth = 2, insert at the head of the singly linked list in the chained hash table and at the node before the tail in the doubly linked list\\}
        
        \State \parbox[t]{\dimexpr\linewidth-\algorithmicindent-\algorithmicindent\relax}{Remove both occurrences of c from input data. Modify the bit flag for the first occurrence to 1 and add another bit flag at the current position of c with a value of 1\\}

    \Else
        \State \parbox[t]{\dimexpr\linewidth-\algorithmicindent-\algorithmicindent\relax} {Increment the depth value in c’s node by 1}
        
        \\
       \State \parbox[t]{\dimexpr\linewidth-\algorithmicindent-\algorithmicindent\relax}{Remove the latest occurrence of c from the input data and add a bit flag of value 1 at the current position}
       
    \EndIf

\Else{
    \State \parbox[t]{\dimexpr\linewidth-\algorithmicindent-\algorithmicindent\relax}{ Append c to buffer\\
	Add a 0-bit flag as a prefix}
	\EndIf}
\EndWhile
\\
\\

// $r_1$ Delta Encoding and writing of compressed characters to //output\\ \\
Traverse the doubly linked list from head to tail

\For {each node}
    \State \parbox[t]{\dimexpr\linewidth-\algorithmicindent-\algorithmicindent\relax}{Output the difference between the current $r_1$ value and greatestr1, followed by the current node’s character and depth value to the compressed output\\
    
    }
     \If{\parbox[t]{\dimexpr\linewidth-\algorithmicindent-\algorithmicindent\relax}{current node's $r_1$ + $r_2$ > greatestr1 + greatestr2}
     \State \parbox[t]{\dimexpr\linewidth-\algorithmicindent-\algorithmicindent\relax}{ greatestr1 = current node's $r_1$\\ greatestr2 = current node's $r_2$\\}
     \EndIf
     }
\EndFor
\\

// Removal of redundant entries \\
\end{algorithmic}

This leads to two separate files/strings – one with compressed characters, hereafter referred to as coutput and another with uncompressed characters, hereafter referred to as uoutput.\\

While compressing large files, the hash table will be extremely large and memory intensive. To reduce the RAM usage, characters present in the hash table for which no repetition has occurred for the previous two $r$ values relative to the $r$ value of the substring currently being processed can be written to file and their nodes can be deleted. Thus, RAM can be freed up while compression is taking place. However, this optimization is beyond the scope of this paper and has not been reflected in the pseudocode.\\

\subsubsection{Decoding}

\begin{algorithmic}[lines]
\State Read coutput data into a matrix of 3 rows wherein:
\State 1st row – $r_1$ values (delta encoded)
\State 2nd row – characters
\State 3rd row – depth value\\

\State Undo delta encoding of $r_1$ values by adding each delta coded value to the $r_1$ value of the previous entry which is uncoded\\

\State int r = 0
\State int prevpos = 0
\While{not end of uoutput file}
\State Read next character in uoutput

\NoThen
\If {bit flag prefix = 0}
		\State \parbox[t]{\dimexpr\linewidth-\algorithmicindent-\algorithmicindent\relax}{ Remove bit flag from data}

\NoThen
\ElsIf {bit flag prefix = 1}
    \State \parbox[t]{\dimexpr\linewidth-\algorithmicindent-\algorithmicindent\relax}{ Search matrix for next character entry for which \\ $r_1\le r-1\le r_1 + depth$ from the prepos + $1^{th}$ \\index onwards}
    
    \\
    \NoThen
    \If {not found}
        \State r = r + 1
        \State \parbox[t]{\dimexpr\linewidth-\algorithmicindent-\algorithmicindent\relax}{Search matrix for character entry for which \\ $r_1\le r-1\le r_1 + depth$}\\
   \EndIf
    
    \State depth = depth – 1
	\State $r_1 = r_1 + 1$
	\State Replace bitflag from uoutput with the character found\\
	
	\NoThen
	\If{depth = 0}
	    \State Remove entry from matrix
	    
        \State prevpos = matrix column index – 1
        
    \Else
        \State prevpos = matrix column index
    \EndIf
\EndIf
\EndWhile

\end{algorithmic}

As observed, the decoding process is significantly simpler and hence runs faster than the encoding process.

\section{EXPERIMENTAL RESULTS}
\begin{figure}[h!]
    \centering
    \includegraphics[scale=0.55]{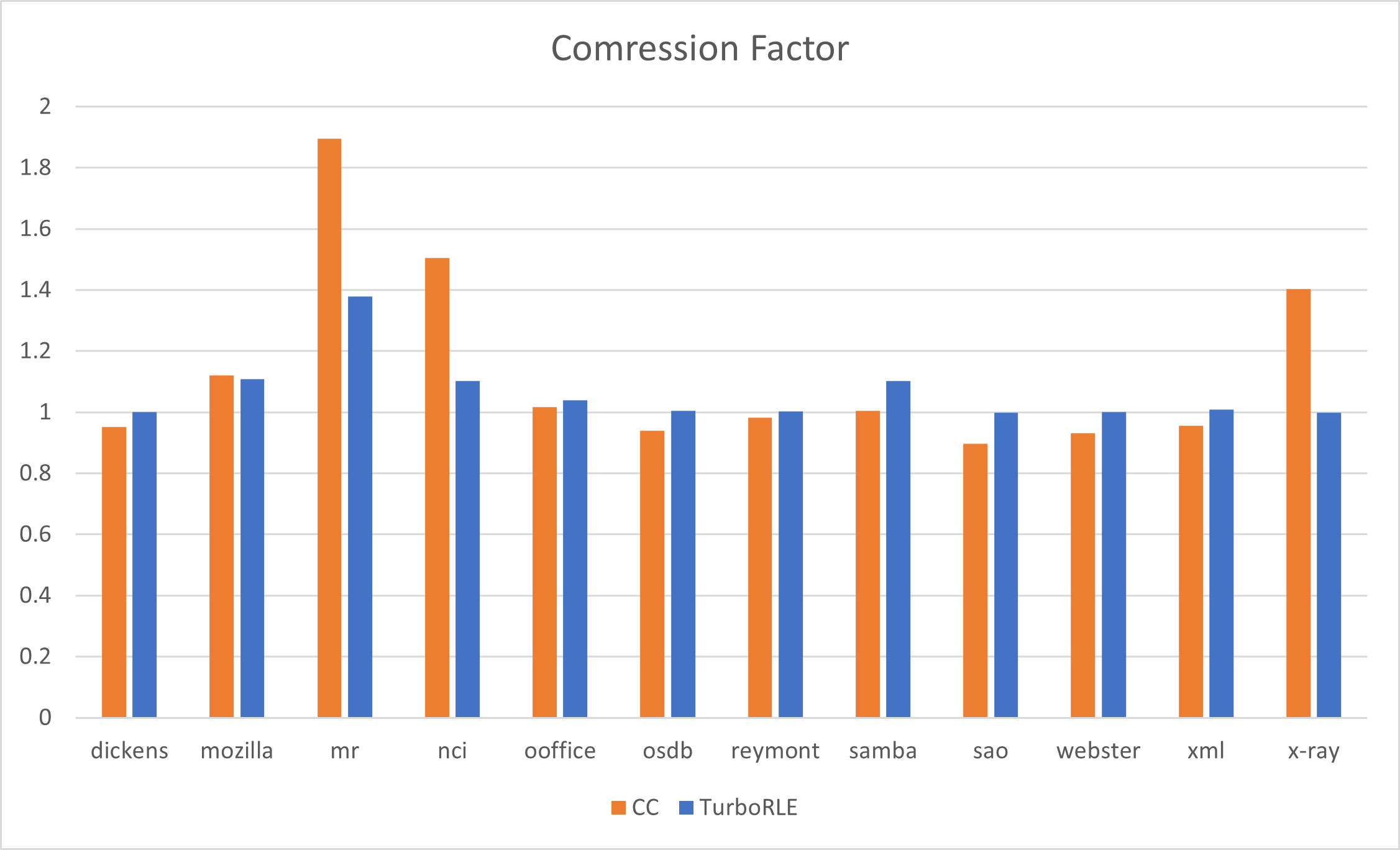}
    \caption{Compression Factor of CC and TurboRLE}
    \label{performance}
\end{figure}

The TurboRLE algorithm and CC were tested on the Silesia Compression Corpus. Figure $\ref{performance}$ shows the algorithms' compression factors.

The overall compression factor achieved by TurboRLE and CC are 1.07 and 1.10 respectively. Hence, CC's compression factor is 2.8\% greater than TurboRLE's. 
\\
\\
It is observed that CC and TurboRLE have similar compression factors for most files. However, for files such as: mr, nci and x-ray, CC outperforms TurboRLE by a significant margin. Since nci is a chemical database following a specific format, mr and x-ray mainly contain black and white pixels, there exists regular repetition of characters in these files i.e. positional redundancy. This validates CC's exploitation of positional redundancy.  
\\
\\
The requirement of bitflag data automatically causes a 1/8 overhead of the input data's size prior to compression. To further improve the compression factor, optimizing this overhead is imperative. Currently, run-time analysis has not been done because the algorithm requires architectural improvement.\\

\section{Conclusion}
In this work, a lossless compression algorithm capable of exploiting positional redundancy has been presented. 

Having tested it on the Silesia Compression Corpus, it is evident that CC outperforms TurboRLE on data containing positional redundancy. By transforming data into the proposed two-dimensional representation, CC compresses characters that are not necessarily contiguous. Hence, it is a generalized form of run length encoding. 
\\

\section*{References}

\begin{enumerate}
\renewcommand*\labelenumi{[\theenumi]}
    \item T. A. Welch. 1984. A Technique for High-Performance Data Compression. Computer 17, 6 (June 1984), 8-19. DOI: https://doi.org/10.1109/MC.1984.1659158
    \item Huffman, D.A.: A method for the construction of minimum-redundancy codes. Proceedings of the Institute of Radio Engineers 40(9), 1098–1101 (1952)
    \item Technical Note TN1023: Understanding PackBits, web.archive.org/web/20080705155158/developer.apple.com/ \newline technotes/tn/tn1023.html.
    
    \item S. W. Smith. The scientist and engineer’s guide to digital signal processing. California Technical Publishing, (1997)

    \item Giovanni Manzini. 2001. An analysis of the Burrows—Wheeler transform. J. ACM 48, 3 (May 2001), 407-430. DOI: https://doi.org/10.1145/382780.382782
    \item Gagie, Travis, and Giovanni Manzini. "Move-to-front, distance coding, and inversion frequencies revisited." Annual Symposium on Combinatorial Pattern Matching. Springer, Berlin, Heidelberg, 2007.
    \item VidyaSagar, M., and JS Rose Victor. "Modified Run Length Encoding Scheme for High Data Compression Rate." International Journal of Advanced Research in Computer Engineering \& Technology (IJARCET) Vol 2 (2013).
    \item David Salomon. 2002. Data compression. In Handbook of massive data sets, James Abello, Panos M. Pardalos, and Mauricio G. C. Resende (Eds.). Kluwer Academic Publishers, Norwell, MA, USA 245-309.

\end{enumerate}

\end{document}